\documentclass[prb,preprint,showpacs]{revtex4}

\usepackage{amsmath,amssymb}

\newcommand{\gammat}{\tilde{\gamma}}

\usepackage{graphicx}

\begin{document}

\widetext

\title{Role of Quantum Coherence and Energetic Disorder on Exciton Transport in Polymer Films}

\author{William Barford$^*$ and Christopher D. P. Duffy}

\affiliation{
Department of Physics and Astronomy, University of Sheffield, Sheffield, S3 7RH, United Kingdom
}

\begin{abstract}
The cross-over from coherent to incoherent exciton transport in disordered polymer films is studied by computationally solving a modified form of the Redfield equation for the exciton density matrix.
This theory models quantum mechanical (ballistic) and incoherent (diffusive) transport as limiting cases. It also reproduces F\"orster transport for certain parameter regimes.
Using model parameters appropriate to polymer thin films it is shown that short-time quantum mechanical coherence increases the exciton diffusion length. It also causes rapid initial energy relaxation and larger line widths. The route to equilibrium is, however, more questionable, as the equilibrium populations of the model do not satisfy the Boltzmann distributions over the site energies. The  Redfield equation for the dimer is solved exactly to provide insight into the numerical results.
\end{abstract}


\maketitle

\section{Introduction}

Energy transport via exciton migration in disordered organic semiconducting solids has
been extensively investigated by theoretical and computational
modelling\cite{mov1,parson,mov2,mov3,meskers}. It is generally assumed that the dephasing rates in
these systems are large compared to the exciton transfer rates, so
that the quantum mechanical phase memory of the exciton is rapidly
lost, rendering the exciton motion incoherent. 
The effective medium approximation was developed by Movaghar and co-workers\cite{mov1,mov2} in the incoherent limit to investigate the role of spatial and energetic disorder on charge and energy transport properties. This mean-field theory agrees remarkably well with Monte Carlo simulations at high enough temperatures, predicting a roughly $-\log(t)$ time dependence for the energy relaxation. At low temperatures, however, mean-field theory overestimates the relaxation pathways\cite{mov3}, and thus fails to predict the `freezing-in' of the energy relaxation observed both experimentally and in Monte Carlo simulations. For a Gaussian distribution of energetic disorder the pseudo-equilibrium diffusion coefficient, $D(T)$,  is found to behave as,
\begin{equation}\label{Eq:50}
    D(T) = D_0 \exp(-(T_0/T)^2)
\end{equation}
at high temperatures (where $T_0$ is proportional to the width of the Gaussian distribution), while at low temperatures it is exponentially activated\cite{mov1}.

Conversely, it is
generally assumed that dephasing rates in biological light
harvesting complexes are generally small enough that the exciton
motion remains coherent for times long enough for the exciton to
successfully reach the reaction center before it can radiatively
recombine. Indeed, in light harvesting complexes the ballistic nature of exciton transport coupled to an energy landscape that `funnels' the exciton to the reaction center leads to particularly efficient energy transport.\cite{amerongen, grondelle} In a disordered energy landscape with sufficiently large energetic disorder, on the other hand, localization occurs in the coherent regime\cite{anderson}.

Dephasing rates in typical organic compounds are
significantly enhanced by intra and inter molecular disorder that scatters the exciton wavefunction\cite{bassler06}. Thus, typical dephasing times are
roughly 1 ps in an ordered conjugated polymer chain\cite{dubin06},
compared to roughly 100 fs in a disordered system\cite{milota04}.
Nonetheless, the advent of ultra-fast spectroscopy allows
relaxation phenomena occurring on sub-100 fs time-scales to be
observed, implying that theoretical modelling of exciton
transport should also take into account quantum coherence effects on these
time-scales. Indeed, more recent Monte Carlo simulations of exciton transport in the incoherent limit fail to predict the observed very fast relaxation processes that lead to a rapid reduction of the average energy and a broadening of the spectral lines\cite{meskers}. Although this rapid energy reduction was attributed to vibrational relaxation\cite{meskers}, as will be demonstrated in this paper, another possible mechanism is coherent energy transport. The `anomalous' red-shifted emission also emphasizes the role of quantum coherence on exciton dynamics, as this has been attributed to the recombination of an exciton delocalized over a pair of polymer chains\cite{meskers}.

As for light harvesting complexes, the successful operation of a polymer photovoltaic device requires
that a photo-excited exciton migrates  to a polymer heterojunction
and then disassociates before it can recombine. An understanding of the role of quantum coherence and energetic disorder on exciton transport is therefore essential if realistic simulations of device properties are to be performed.

This paper presents computational studies of the cross-over from coherent to
incoherent exciton transport. 
The model that we adopt  assumes a localized basis whereby an exciton initially created on a chromophore migrates to neighboring chromophores. This model retains the properties of fully coherent and fully incoherent transport as limiting cases. In particular, for short times the quantum mechanical processes lead to a coherent wavepacket that is delocalized over more than one chromophore. At long times dephasing processes destroy phase coherence, and the motion becomes `classical'. F\"orster-like dipole-dipole induced exciton transport is described by a limiting case of this  model. We show that short-time coherent motion leads to increased exciton diffusion lengths, and to initial rapid energy relaxation and increased line widths.

The plan of this paper is as follows. The next section outlines the model, while Section III describes the numerical techniques to solve it. Section IV describes and discusses our results, and we conclude in Section V.

\section{The Model}\label{Se:2}

Exciton transfer from chromophore $m$ to chromophore $n$, mediated by the dipole-dipole interaction, is parameterized by the energy transfer integral, $J'_{mn}$. Assuming that each chromophore exists either in its ground state or a single excited state, and that the excitation energy of the $m$th chromophore is $\epsilon_m$, the Hamiltonian that describes the coherent exciton dynamics in the absence of exciton interactions is,
\begin{equation}\label{Eq:1}
H = \sum_m \epsilon_m |m\rangle \langle m | +
\sum_{mn}\left(J'_{mn}|m\rangle \langle n | + J'_{nm}|n\rangle
\langle m |\right).
\end{equation}
The basis state $|m\rangle$ represents an exciton localized on the $m$th chromophore. In general, exciton interactions, particularly with defects and the heat bath via the exchange of phonons, will cause a loss of phase coherence.
This behavior
is conveniently modelled by an equation of motion for the reduced
density operator, $\hat{\rho}$, defined by
\begin{equation}\label{Eq:2}
\hat{\rho} = Tr\{\hat{W}(t)\},
\end{equation}
where $\hat{W}(t)$ is the full density operator and the trace is
over all the degrees of freedom of the environment.

In a localized exciton basis the matrix elements of the reduced
density operator are $\rho_{mn} = \langle m | \hat{\rho} |n \rangle$.
An equation of motion for these
matrix elements  may be formally derived by performing the trace in Eq.\ (\ref{Eq:2})\cite{may}.
In this paper, however, we assume a semi-phenomenological approach and adopt
a Redfield-like equation that describes coherent  and incoherent
processes as limiting cases, and in general models the cross-over
from short-time coherent behavior to long-time incoherent
behavior. This model is defined as follows by decomposing the
equation of motion for the matrix elements into its constituent
parts:
\begin{equation}\label{Eq:3}
 \frac{\partial \rho_{mn}(t)}{\partial t} = L_1\rho_{mn}(t) + L_2\rho_{mn}(t) +
 L_3\rho_{mn}(t) + L_4\rho_{mn}(t),
\end{equation}
where
\begin{eqnarray}\label{Eq:4}
L_1\rho_{mn} =  -i\left[ H, \rho\right]_{mn} = - i\omega_{mn}\rho_{mn} - \frac{i}{\hbar}
\sum_{\ell}\left( J_{m\ell}\rho_{\ell n} - J_{\ell n} \rho_{m
\ell} \right),
\end{eqnarray}
\begin{eqnarray}\label{Eq:5}
L_2\rho_{mn} = - \delta_{mn} \sum_{\ell}\left( k_{m \ell} \rho_{mm} -
k_{\ell m} \rho_{\ell \ell}\right),
\end{eqnarray}
\begin{eqnarray}\label{Eq:6}
L_3\rho_{mn} = - 2\tilde{\gamma}_{mn}(1-\delta_{mn})\rho_{mn},
\end{eqnarray}
and 
\begin{eqnarray}\label{Eq:7}
L_4\rho_{mn} = - \delta_{mn}\rho_{mn}/\tau_m.
\end{eqnarray}
In these equations we have defined, $J_{mn} = J'_{mn}/\hbar$, $\omega_{mn} = (\epsilon_m - \epsilon_n)/\hbar$,
$2\tilde{\gamma}_{mn} = \gamma_m + \gamma_n + 2\gamma_0$,  and
$\gamma_m = \frac{1}{2}\sum_{m \rightarrow \ell} k_{ml}$.

$L_1\rho$ represents the coherent, ballistic motion of the
exciton, and describes the exciton motion in accordance with the
time-dependent Schr\"odinger equation,
\begin{equation}\label{Eq:8}
i\hbar \frac{\textrm{d}|\Psi(t)\rangle}{\textrm{d}t} =
H|\Psi(t)\rangle,
\end{equation}
where $H$ is given by Eq.\ (\ref{Eq:1}).

$L_2\rho$ represents the incoherent, diffusive motion of the
exciton arising from population transfer from chromophore to chromophore.
Formally, this is associated with vibrational-induced exciton
transfer via the spatial modulation of the dipole-dipole
coupling\cite{may}. With this term alone Eq.\ (\ref{Eq:3}) is equivalent to
the Pauli master equation,
\begin{equation}\label{Eq:9}
 \frac{\partial P_{m}(t)}{\partial t} = - \sum_{\ell}\left( k_{m \ell} P_m -
k_{\ell m} P_{\ell} \right),
\end{equation}
where $P_m \equiv \rho_{mm}$. Energy relaxation occurs when the rate for energy transfer to a higher energy chromophore is smaller than the rate for energy transfer to a lower energy chromophore.
It is customary to assume that the rates satisfy,
\begin{equation}\label{Eq:10}
\frac{k_{m \ell }}{k_{\ell m}} = \exp(-\hbar \omega_{\ell m}/k_BT),
\end{equation}
which guarantees that the equilibrium populations satisfy the Boltzmann distribution for Eq.\ (\ref{Eq:9}). As will be shown later, however,  Eq.\ (\ref{Eq:10}) does not guarantee that the equilibrium populations of the Redfield equation (Eq.\ (\ref{Eq:3})) satisfy the Boltzmann distribution.

$L_3\rho$ represents the coherent (or transverse) dephasing of the
off-diagonal elements of the density matrix. In the Bloch model
the transverse dephasing time, $T_2 = (2\tilde{\gamma})^{-1}$.

Finally, $L_4\rho$ represents the population decay via
exciton recombination mechanisms. If the dominant decay is via radiative
recombination then,
\begin{equation}\label{Eq:11}
\tau_m = \tau_0 \left(\frac{\bar{\epsilon}}{\epsilon_m} \right)^3,
\end{equation}
where $\bar{\epsilon}$ is the average chromophore energy.

Eq.\ (\ref{Eq:3})  has been extensively studied for translationally invariant systems (see \cite{reineker}, \cite{silinsh} or \cite{amerongen}, for example).
Exact results for the mean-square-displacement, $\langle r^2
\rangle$, have been obtained  by
Reineker\cite{reineker} for a
$d$-dimensional cubic hyper-lattice. For an exciton created at the origin at time $t=0$,
\begin{equation}\label{Eq:12}
    \langle r^2(t) \rangle = a^2 d \left( 2 \left(k + \frac{J^2}{\tilde{\gamma}}\right)t
    + \frac{J^2}{\tilde{\gamma}^2}\left( \exp(-2\tilde{\gamma}t) - 1 \right)\right),
\end{equation}
where $a$ is the lattice parameter. For long times ($t \gg \gammat^{-1}$) this expression reproduces the random walk result,
\begin{equation}\label{Eq:13}
    \langle r^2 \rangle = 2 a^2 d \left(k + \frac{J^2}{\tilde{\gamma}}\right)t +\textrm{ constant},
\end{equation}
with an additional F\"orster contribution  to the diffusion
(namely, the $J^2/\gammat$ term) arising from the dipole-dipole coupling. This result reflects the fact that Eq.\ (\ref{Eq:3}) approximates to the diffusion equation in the long-time limit\cite{reineker}. Thus, in this model in the long-time limit the dipole-dipole interaction contributes to incoherent motion in two ways. First, explicitly from $L_2\rho_{mn}$ (where $k$ formally is proportional to the spatial derivative of $J_{mn}$\cite{may}) and second, implicitly from the F\"orster mechanism.

For short
times ($t \rightarrow 0$), the expected ballistic result
is modified by an anomalous  linear term resulting from the
assumptions in the derivation of Eq.\ (\ref{Eq:12})\cite{reineker}:
\begin{equation}\label{Eq:14}
    \langle r^2 \rangle = 2a^2 d \left( kt + J^2t^2 \right).
\end{equation}
It is immediately apparent from Eq.\ (\ref{Eq:12}) that for short times (but
for $t > k/J^2$) the exciton transport  is dominated by coherent,
ballistic processes, and therefore the exciton travels much
further than it would if only incoherent, diffusive processes are
considered.

Finally, the fully quantum mechanical behavior of Eq.\ (\ref{Eq:8}) is also reproduced from Eq.\ (\ref{Eq:12}) by setting  $k = \gammat = 0$,
\begin{equation}\label{Eq:15}
    \langle r^2 \rangle = 2a^2 d J^2t^2.
\end{equation}

Appendix \ref{Appendix:B} describes the solution of Eq.\ (\ref{Eq:3}) for the special case of a dimer.
For equal site energies this solution illustrates the decay of coherence and the F\"orster limit
when $\gamma \gg J$, while for unequal site energies it illustrates the persistence of coherence 
 and that the equilibrium populations are not determined by the Boltzmann distribution over the site energies when the rates satisfy Eq. (\ref{Eq:10}).
To our knowledge no analytical results exist for the transport properties associated with Eq.\ (\ref{Eq:3}) in a general disordered energy landscape. The numerical techniques to solve this model are therefore described in the next section.

\section{Methodology}\label{Se:2}

Eq.\ (\ref{Eq:3}) may be expressed as the set of $M \times M$ simultaneous
equations,
\begin{equation}\label{Eq:16}
 \frac{dy_i(t)}{dt} = \sum_j A_{ij}y_j(t),
\end{equation}
where $M = N_{\textrm{sites}}^2$ and $N_{\textrm{sites}}$
is the number of sites in the lattice,
$y_i(t) \equiv \rho_{mn}(t)$, and $\textbf{A}$ is the
coupling matrix. The formal solution of Eq.\ (\ref{Eq:16}) is,
\begin{equation}\label{Eq:17}
 y_i(t) = \sum_{jk} S_{ij} \exp(\lambda_j t) S_{jk}^{-1} y_k(0).
\end{equation}
Here, $\textbf{S}$ is the matrix whose columns are the
eigenvectors of $\textbf{A}$, $\{\lambda\}$ are  the eigenvalues
of $\textbf{A}$, and $\{y(0)\}$ are the initial conditions.
Assuming that the $m$th chromophore is excited at $t=0$, we
have,
\begin{equation}\label{Eq:18}
 y_i(t) = \sum_{j} S_{ij} \exp(\lambda_j t) S_{jm}^{-1}.
\end{equation}

The evaluation of  Eq.\ (\ref{Eq:18}) requires a diagonalization of
the $M \times M$ sparse, complex, and non-hermitian transformation matrix, $\textbf{A}$.
 For most problems of
interest this matrix is far too large to numerically diagonalize
completely. However, the real parts of the eigenvalues of
$\textbf{A}$ are the rates for the decay of the eigenmodes.
Consequently, for times greater than an arbitrary cut-off, $\tau$,
only eigenvalues whose real parts satisfy $\Re|\lambda_i| <
\tau^{-1}$ need to be computed. If only this long-time behavior is
required, it is only necessary to diagonalize
$\textbf{A}$ for a small sub-set of the entire spectrum using
sparse-matrix diagonalization techniques\cite{NAG2}. This approach is further described in Appendix \ref{Appendix:A}. The long-time behavior can then be matched to the short-time behavior obtained by standard numerical integration techniques, as described below.

In practice, the equation of motion for the density matrix, Eq.\ (\ref{Eq:3}), may be
solved more efficiently for large systems by standard numerical  time-discretization techniques. To
ensure both stability and accuracy the Crank-Nicolson scheme -- an
average of implicit and explicit forward-time-centered-space
discretization methods -- is employed\cite{num press2, NAG3}. In
order to adapt the Crank-Nicolson scheme for Eq.\ (\ref{Eq:3}) it is
necessary to adopt the Operator Splitting Method\cite{num press3}
by decomposing the spatial differential operator on the
right-hand-side of Eq.\ (\ref{Eq:3}) into a sum of zero or one-dimensional
operators.

Expectation values of operators corresponding to dynamical variables are found from the density matrix in the usual manner, via,
\begin{equation}\label{Eq:19}
    \langle \hat{O}(t) \rangle = \sum_{mn}   \rho_{mn}(t)O_{nm}.
\end{equation}

\section{Results and Discussion}

For an $N_{\textrm{site}}$ site lattice the number of matrix elements (or components) of Eq.\ (\ref{Eq:3}) is $N_{\textrm{site}}^2$, rendering this a computationally very expensive problem. To obtain numerical results from time-scales of $10^{-15}$ to $10^{-9}$ seconds it has therefore been necessary to restrict the size and dimensionality of the lattice. In this paper we describe results for  two-dimensional square lattices of up to $41 \times 41$ sites. 

Our parametrization of the model follows closely that of Meskers \textit{et al.}\cite{meskers}. We take nearest neighbor interactions on a square lattice, with $k=0.00088$ eV ($\equiv$ (5 ps)$^{-1}$), $\tau_0 = 300$ ps, $\gamma_0=0.0207$  eV ($\equiv$ (200 fs)$^{-1}$), while $J$ (which models the strength of the quantum coherence) is an independent parameter, varying from $0.02$ eV to $0.00414$ eV ($\equiv$ (1 ps)$^{-1}$)\cite{footnote1}. The value of $J = 0.00414$ eV corresponds to a F\"orster transfer rate of $J^2/\gamma_0 = 0.00088$ eV, so in the limit that $k=0$ and long times the model maps directly onto the Mesker parametrization. The energetic disorder is modelled by a Gaussian distribution function\cite{num press4} of mean energy, $\bar{\epsilon} = 3.0$ eV and $\sigma = 0.07$ eV. We perform averages over ensembles of five realizations of the disorder. The system is initially excited at the origin at $t=0$ with an energy of $3.1$ eV.

\subsection{Root-mean-square distance}

\begin{figure}[tb]
\begin{center}
\includegraphics[scale=0.60]{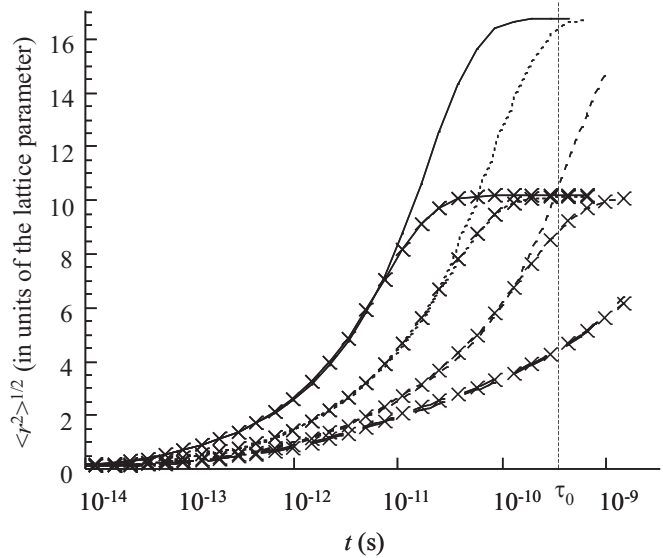}
\end{center}
\caption{The mean root-mean-square distance, $\langle r^2
\rangle^{1/2}$, versus time for various values of $J$ and lattice
size: $J=0.02$ eV (solid curves); $J=0.01$ eV (dotted curves);
$J=0.00414$ eV (short-dashed curves); $J = 0$ (long-dashed
curves); $41\times 41$ lattice (no symbols); $25\times 25$ lattice ($\times$). Also shown is the mean recombination life time, $\tau_0 = 300$ ps.} \label{Fi:1}
\end{figure}

\begin{figure}[tb]
\begin{center}
\includegraphics[scale=0.60]{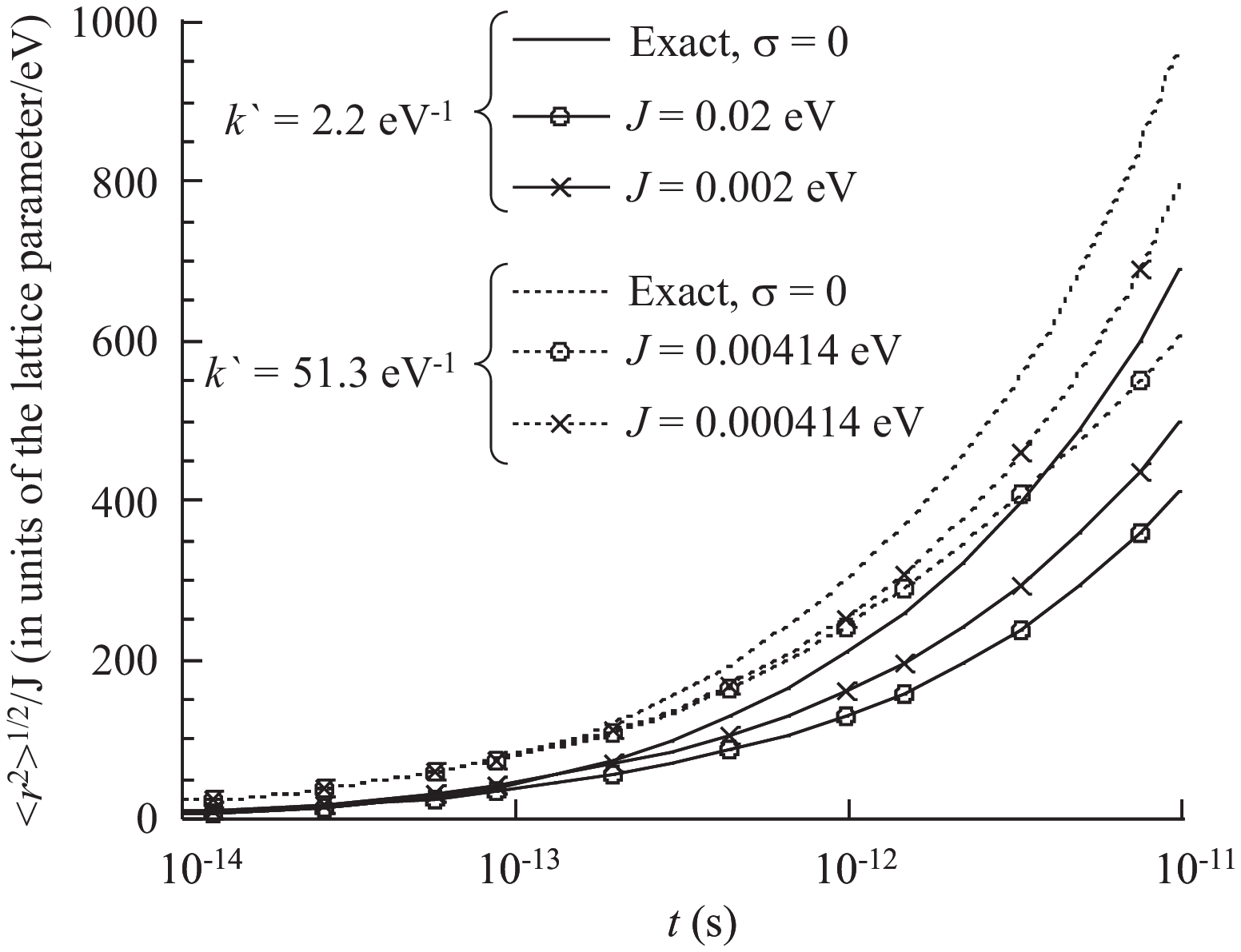}
\end{center}
\caption{Scaled root-mean-square distance, $\langle r^2
\rangle^{1/2}/J$, versus time for various values of energetic disorder and $k' = k/J^2$ on a $41\times 41$ lattice. $k' = 2.2$ eV$^{-1}$ (solid curves); $k' = 51.3$ eV$^{-1}$ (dotted curves); $\sigma = 0.07$ eV (curves with symbols); $\sigma = 0$ (curves with no symbols). The exact results are obtained from Eq.\ (\ref{Eq:12}).}\label{Fi:2}
\end{figure}

We first investigate the role of quantum coherence on the root-mean-square distance, $\langle r^2(t) \rangle^{1/2}$, travelled by the exciton at a time $t$. Fig.\ \ref{Fi:1} shows $\langle r^2 \rangle^{1/2}$ for various values of $J$. As expected, larger values of $J$ lead to greater distances travelled. For a value of $J=0.02$ eV the exciton has reached the boundaries of the $41 \times 41$ square lattice within its mean life time, $\tau_0 = 300$ ps. In contrast, in the classical limit ($J=0$) the exciton has only travelled ca.\ $4$ lattice units in this time.

We next discuss the role of energetic disorder on the root-mean-square distance. As Eq.\ (\ref{Eq:50}) indicates, in the incoherent limit the scaled pseudo-equilibrium constant, $D/D_0$, is a function of disorder (and temperature) only. In contrast, for purely coherent motion diffusion is related to both the exciton band width and the scale of the disorder, and indeed localization occurs for sufficiently larger disorder\cite{anderson}.
Eq. (\ref{Eq:12}) indicates that $\langle r^2 \rangle^{1/2}/J$ is a function of $\gamma$ and $k/J^2 \equiv k'$. Thus, in the absence of disorder plots of $\langle r^2 \rangle^{1/2}/J$ versus time for fixed values of $\gamma$ and $k'$ will coincide for all values of $J$. Fig.\ \ref{Fi:2} shows that, as expected, disorder reduces the value of $\langle r^2 \rangle^{1/2}$. Rather unexpectedly the scaled root-mean-square distance increases with decreasing $J$. This is an artefact, however, of the fact that disorder is more effective at hindering transport when quasi-equilibrium is reached, which takes longer to achieve for smaller values of $J$. A full analysis of the scaling  of diffusion with disorder requires further more extensive numerical calculations and analytical studies.

\subsection{Energy relaxation}

\begin{figure}[tb]
\begin{center}
\includegraphics[scale=0.60]{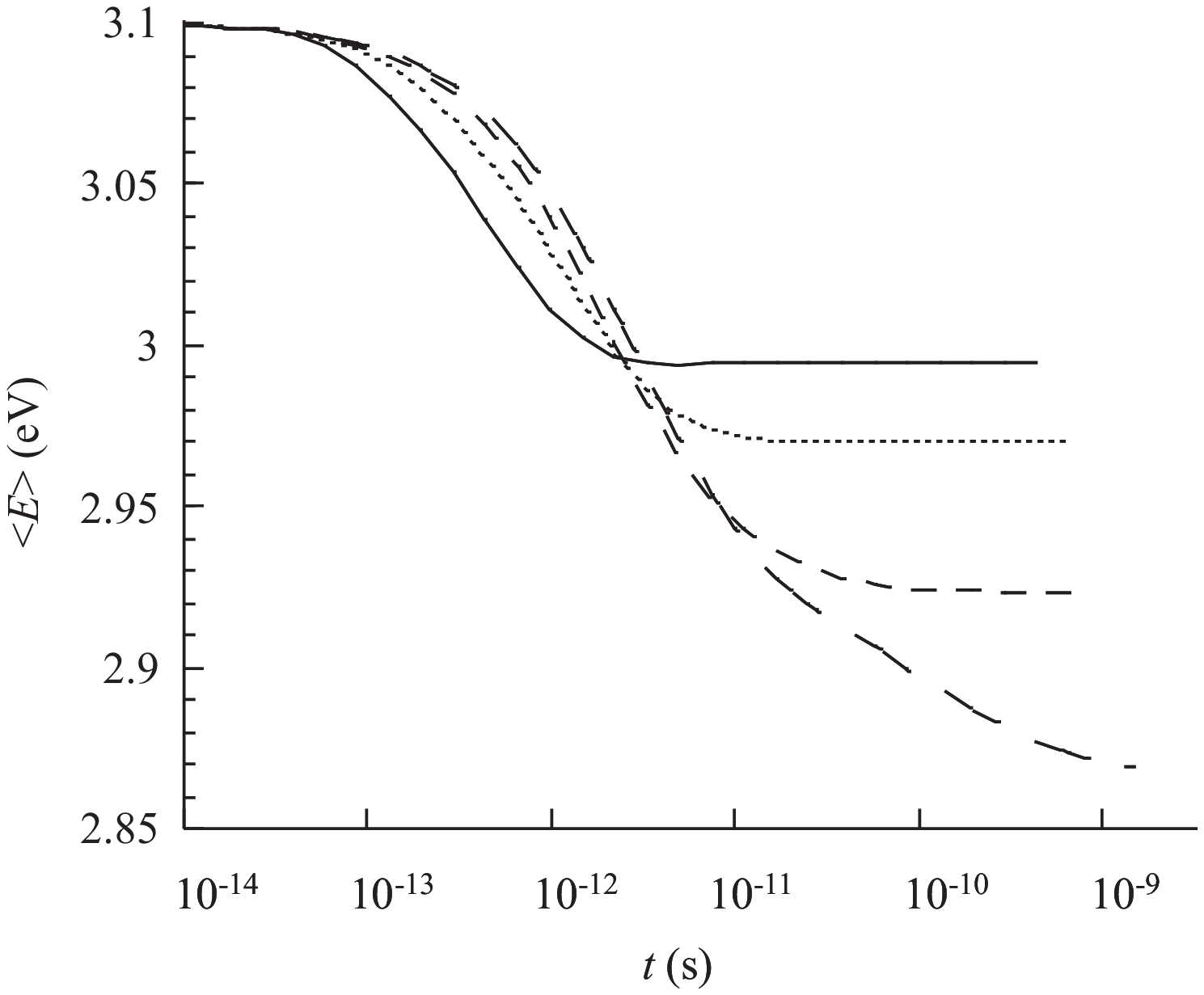}
\end{center}
\caption{The mean energy, $\langle E \rangle$, versus time for
various values of $J$ for the $41\times 41$ lattice: $J=0.02$ eV
(solid curves); $J=0.01$ eV (dotted curves); $J=0.00414$ eV
(short-dashed curves); $J = 0$ (long-dashed curves). }
\label{Fi:4}
\end{figure}

The expectation value of the energy as a function of time is obtained via Eq.\ (\ref{Eq:19}) with $\hat{O} \equiv \hat{H}$, and $\hat{H}$ given by Eq.\ (\ref{Eq:1}). As shown in Fig.\ \ref{Fi:4}, the initial rate of energy relaxation increases as $J$ increases. By $10^{-12}$ s the energy has relaxed by $0.1$ eV when $J=0.02$ eV and by $0.05$ eV in the classical limit ($J=0$). This result can be understood by the observation that in the coherent regime the exciton initially forms a wavepacket that travels ballistically, and hence rapidly samples a wide ensemble of sites.

Notice, however, that the equilibrium value of $\langle E \rangle$ \emph{increases} with increasing $J$, and for non-zero $J$ it obviously does not satisfy the canonical ensemble value given by the classical (Boltzmann) distribution over the site energies $\{ \epsilon_m\}$. This result is a consequence of the fact that for this model with a disordered energy landscape the equilibrium values of the coherences (i.e.\ the off-diagonal density matrix elements) are not identically zero. Thus, because of quantum mechanical delocalization, the probability for occupying a high energy site is higher than that predicted by the Boltzmann distribution. This increase in the potential energy caused by the delocalization onto higher energy sites is not entirely compensated by the kinetic energy reduction arising from the quantum mechanical delocalization. 
A full treatment of a dimer with unequal energies is given in Appendix \ref{Appendix:B} to further illustrate this point.

\begin{figure}[tb]
\begin{center}
\includegraphics[scale=0.60]{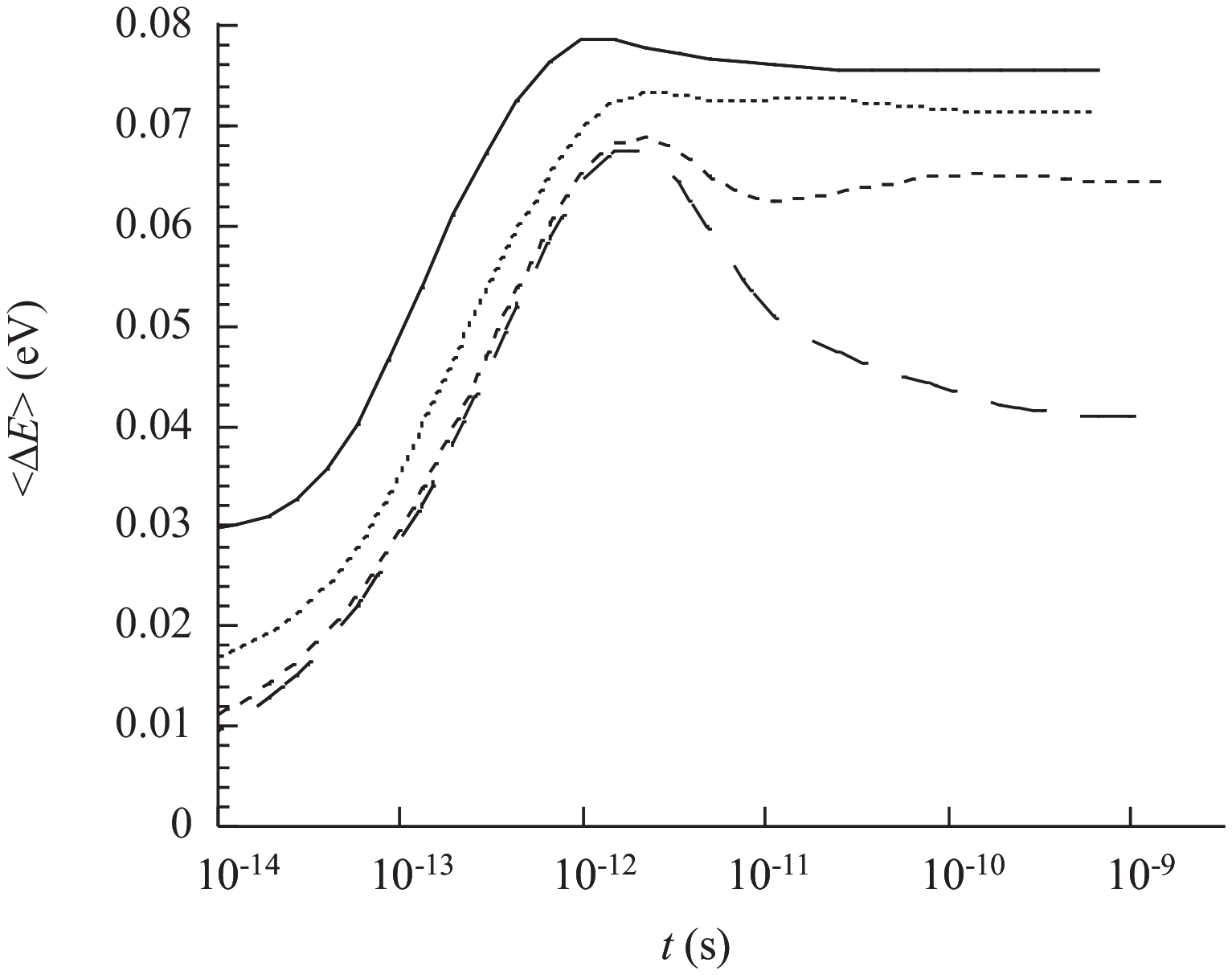}
\end{center}
\caption{The mean standard deviation of the energy, $\langle
\Delta E \rangle$, versus time for various values of $J$ for the
$25\times 25$ lattice: $J=0.02$ eV (solid curves); $J=0.01$ eV
(dotted curves); $J=0.00414$ eV (short-dashed curves); $J = 0$
(long-dashed curves).} \label{Fi:5}
\end{figure}

Fig.\ \ref{Fi:5} shows the line width, $\langle \Delta E \rangle = \langle  E^2 - \langle E \rangle^2 \rangle^{1/2}$, for various values of $J$. As expected from the previous discussion concerning the initial faster delocalization of the exciton in the quantum mechanical limit, the line width increases more rapidly for larger values of $J$. The equilibrium line widths are also larger for larger values of $J$, because the measurement of energy is with respect to the Hamiltonian, Eq.\ (\ref{Eq:1}), whereas the system is not in an eigenstate of that Hamiltonian.

\section{Conclusions}

The successful operation of a polymer photovoltaic device requires
that a photo-excited exciton travels to a polymer heterojunction
and then disassociates before it can recombine. However, exciton
transport in molecular systems is strongly dependent on molecular
order, as this both determines the exciton dephasing times and
the energy landscape through which the exciton travels.

This paper describes numerical solutions of a modified Redfield equation
for the exciton reduced density matrix in order to study the role of
quantum mechanical coherence and energetic disorder both on the exciton diffusion length and its energetic relaxation. Using model parameters appropriate to polymer thin films, we showed that increased quantum coherence (achieved by increasing the exciton band width) leads to increased exciton diffusion lengths. It also leads to initially more rapid energy relaxation and to wider line widths, in qualitative agreement with experiment\cite{meskers}.
Quite generally, increased disorder implies shorter coherence times and energetic trapping, thereby strongly inhibiting the exciton's ability to migrate successfully to a polymer heterojunction.

Although the model is appropriate for the short-time transport and energy relaxation processes, its applicability for describing the long-time route to equilibrium is more questionable. In particular, we showed that the particular assumption of Eq.\ (\ref{Eq:10}) for the rates for population transfer do not in equilibrium reproduce the Boltzmann distribution over the site energies of Eq.\ (\ref{Eq:1}). (This result was proved for the special case of the dimer.) The cause of this discrepancy is the choice of basis for the exciton transport. As described in ref\cite{may}, taking as a basis the exciton eigenstates of Eq.\ (\ref{Eq:1}) would give equilibrium described by the Boltzmann distribution \emph{over these eigenstates}. Unfortunately, however, this prescription does not correspond to the physically intuitive picture of local exciton transfer from site to site (and consequently does not reproduce F\"orster transfer as a limiting case).

Further work will include  simulations over larger lattices in three dimensions and the possible development of a mean field theory in order to understand the relations between disorder and quasi-equilibrium diffusion.

\appendix

\section{Exact Solution of the Redfield Equation for the Dimer}\label{Appendix:B}

The exact solution of the Redfield equation of motion of the density matrix for the dimer illustrates a number of important
results.  For equal site energies it illustrates the decay of coherence and the F\"orster limit
when $\gamma \gg J$. For unequal site energies it illustrates the persistence of coherence 
 and that the equilibrium populations are not determined by the Boltzmann distribution over the site energies when the rates satisfy the  expression, Eq.\ (\ref{Eq:10}).

The equation of motion of the matrix elements for the dimer in the absence of population loss are,
\begin{equation}\label{}
   \frac{\partial \rho_{11}}{\partial t} = -iJ(\rho_{21}-\rho_{12}) -k_{12}\rho_{11} + k_{21}\rho_{22},
\end{equation}
\begin{equation}\label{}
   \frac{\partial \rho_{12}}{\partial t} = -i\omega\rho_{12} -iJ(\rho_{22}-\rho_{11})-2\gamma_0\rho_{12},
\end{equation}
\begin{equation}\label{}
   \frac{\partial \rho_{22}}{\partial t} = iJ(\rho_{21}-\rho_{12}) + k_{12}\rho_{11} - k_{21}\rho_{22},
\end{equation}
and
\begin{equation}\label{}
   \frac{\partial \rho_{21}}{\partial t} = i\omega\rho_{21} + iJ(\rho_{22}-\rho_{11})-2\gamma_0\rho_{12},
\end{equation}
where $\omega = (\epsilon_1 -\epsilon_2)/\hbar$ and we have set $\tilde{\gamma} \equiv \gamma_0$. The initial conditions are taken as $\rho_{11}(0) = 1$ and $ \rho_{22}(0) = \rho_{12}(0) = \rho_{21}(0) = 0$.

When the site energies are equal $k_{12} = k_{21} \equiv k$, $\omega = 0$, and the equations are easily solved 
by Laplace transforms to give,
\begin{equation}\label{}
P_1(t) \equiv \rho_{11}(t) = \frac{1}{2}\left( 1 + \left(\cos(\Omega t)
+\left(\frac{\gamma_0-k}{\Omega}\right)\sin(\Omega t)\right)\exp(-(k
+ \gamma_0) t)\right),
\end{equation}
\begin{equation}\label{}
P_2(t) \equiv \rho_{22}(t) = 1 -P_1,
\end{equation}
\begin{equation}\label{}
\Re (\rho_{12}(t)) \equiv \Re(\rho_{21}(t)) = 0,
\end{equation}
and
\begin{equation}\label{}
\Im(\rho_{12}(t)) \equiv - \Im(\rho_{21}(t)) = \frac{2J}{\Omega}\sin(\Omega t)\exp(-(k + \gamma_0)t),
\end{equation}
where
\begin{equation}
\Omega = \left( 4 J^2 - (k-\gamma_0)^2\right)^{1/2}.
\end{equation}
Notice that the real components of the coherences are identically zero. Also, for under-damped systems the populations decay to their classical values of $1/2$ and the imaginary components of the coherences decay to zero on a time-scale of $(k + \gamma_0)^{-1}$.

The F\"orster limit is derived for the case that $k=0$ and $\gamma_0
\gg J$. Then,
\begin{equation}\label{Eq:10.48}
P_1(t) \rightarrow \frac{1}{2}\left( 1 + \exp(-2 \tilde{k} t)\right) +
O(J/\gamma_0)^2,
\end{equation}
where $\tilde{k}$, the F\"orster rate, is $\tilde{k} = J^2/\gamma_0$, and $P_1$ is a
solution of the Pauli master equation, Eq.\ (\ref{Eq:9}), with $k_{\ell m}$
replaced by $\tilde{k}$.

When the site energies are unequal the transfer rates are no longer symmetric and the resulting solutions are considerably more complicated. Defining $k_{12} = k$, $k_{21} = k\textrm{e}^{-\Delta}$, and $\Delta = (\epsilon_1 - \epsilon_2)/k_BT$, we obtain,

\begin{eqnarray}\label{}
P_1(t) = \frac{1}{2}\sum_{i}^3 \sum_{j,k(j\ne k \ne i)}^3\frac{(r_i(r_i^2 + Ar_i+B)+C)}{r_i(r_i-r_j)(r_i-r_k)}\exp(r_it) -\frac{C}{r_1r_2r_3},
\end{eqnarray}
\begin{eqnarray}\label{}
P_2(t) = 1 - P_1(t),
\end{eqnarray}
\begin{eqnarray}\label{}
\Im(\rho_{12}(t)) = && -\frac{1}{4J}\sum_{i}^3 \sum_{j,k(j\ne k \ne i)}^3\frac{(r_i(r_i^2 + Ar_i+B)+C)}{(r_i-r_j)(r_i-r_k)}\exp(r_it)
\\ \nonumber
&& -\frac{\alpha}{2J}P_{1}+\frac{k\textrm{e}^{-\Delta}}{2J}
\end{eqnarray}
and
\begin{eqnarray}\label{}
\Re(\rho_{12}(t)) = && \frac{1}{4J\omega}\sum_{i}^3 \sum_{j,k(j\ne k \ne i)}^3\frac{(r_i+\alpha)(r_i(r_i^2 + Ar_i+B)+C)}{(r_i-r_j)(r_i-r_k)}\exp(r_it) 
\\ \nonumber
&& -\frac{2\gamma_0}{\omega}\Im(\rho_{12}) -\frac{J}{\omega}(1-2P_1)
\end{eqnarray}
where,
\begin{eqnarray}\label{}
r_1 = \frac{1}{6}\left( -2a + \frac{2^{4/3}(a^2-3b)}{\xi^{1/3}} + 2^{2/3}\xi^{1/3}\right),
\end{eqnarray}
\begin{eqnarray}\label{}
r_2 = r_3^*  = \frac{1}{12}\left( -4a + \frac{2^{4/3}(1+i\sqrt{3})(a^2-3b)}{\xi^{1/3}} - 2^{2/3}(1-i\sqrt{3})\xi^{1/3}\right),
\end{eqnarray}
\begin{eqnarray}\label{}
\xi = -2a^3 +9ab -27c + 3\sqrt{3}\sqrt{-a^2b^2+4b^3+4a^3c-18abc+27dc^2},
\end{eqnarray}
\begin{equation}\label{}
    a = 4\gamma_0 + \alpha,
\end{equation}
\begin{equation}\label{}
   b = 4J^2 + 4\gamma_0 \alpha +|\lambda|^2,
\end{equation}
\begin{equation}\label{}
   c = 8J^2\gamma_0 +\alpha|\lambda|^2,
\end{equation}
\begin{equation}\label{}
   A = 4\gamma_0 + k\textrm{e}^{-\Delta},
\end{equation}
\begin{equation}\label{}
   B = 4\gamma_0 k\textrm{e}^{-\Delta} + 2J^2 +|\lambda|^2,
\end{equation}
\begin{equation}\label{}
   C = 4\gamma_0 J^2 + k\textrm{e}^{-\Delta}|\lambda|^2,
\end{equation}
\begin{equation}\label{}
    \lambda = i\omega -2\gamma_0,
\end{equation}
and
\begin{equation}\label{}
    \alpha = k(1+\textrm{e}^{-\Delta}).
\end{equation}

\begin{figure}[tb]
\begin{center}
\includegraphics[scale=0.60]{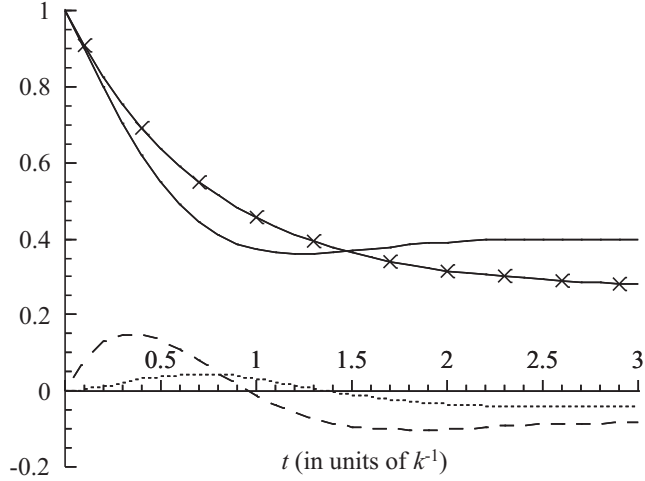}
\end{center}
\caption{$\rho_{11}(t)$ (solid curve), $\Re (\rho_{12}(t))$ (dotted
curve), and $\Im (\rho_{12}(t))$ (dashed curve) versus time for the dimer with
$J=\gamma_0=k=\epsilon_1 = k_B T   =1$, and $\epsilon_2 = 0$. Also
shown is  $\rho_{11}(t)$ (solid curve
with crosses) in the classical limit ($J=0$).}
\label{Fi:6a}
\end{figure}

$\rho_{11}(t)$, $\Re (\rho_{12}(t))$, and $\Im (\rho_{12}(t))$ are plotted in Fig.\ \ref{Fi:6a} for the parameter set $J=\gamma_0=k=\epsilon_1 = k_B T   =1$, and $\epsilon_2 = 0$. For unequal site energies the equilibrium values of the classical populations do not satisfy the Boltzmann distribution and the  coherences are not zero. 

\begin{figure}[tb]
\begin{center}
\includegraphics[scale=0.60]{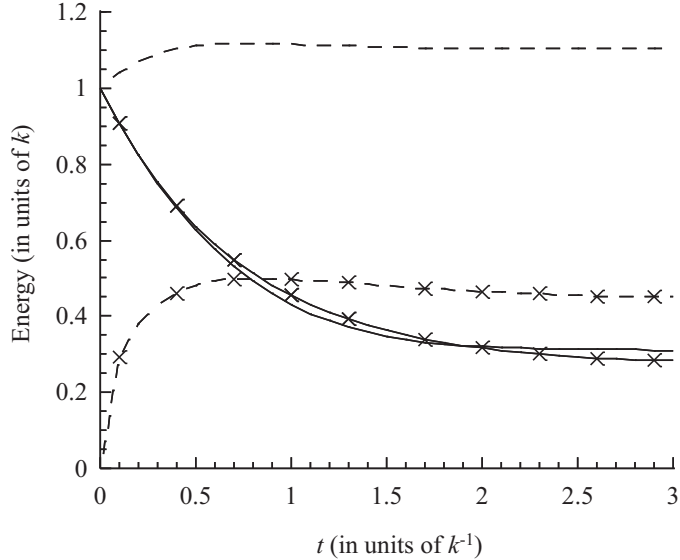}
\end{center}
\caption{$\langle E(t) \rangle$ (solid curves) and $\langle \Delta E(t) \rangle$ (dashed curves) versus time for the dimer with
$J=\gamma_0=k=\epsilon_1 = k_B T   =1$, and $\epsilon_2 = 0$. The curves with crosses are the classical limit ($J=0$). Notice that although
the energy relaxation is initially faster in the mixed regime, the
equilibrium energy exceeds the result of the classical limit. For equal site energies $\langle \Delta E(t) \rangle = |J|$.}
\label{Fi:6b}
\end{figure}

The expectation value of the energy of the dimer is
\begin{eqnarray}\label{Eq:B26}
\langle E(t) \rangle =  \epsilon_1\rho_{11}(t) +  \epsilon_2\rho_{22}(t) + 2J\Re \rho_{12}(t),
\end{eqnarray}
and, as shown by Fig.\ \ref{Fi:6b}, although the initial relaxation is faster than in the classical limit the equilibrium value exceeds the classical value. Quantum mechanical  delocalization onto the higher energy site increases the average energy, because $P_1$ is larger than its Boltzmann value, thus raising the potential energy. This increase in potential energy is partially compensated by a reduction in the kinetic energy arising from the last term in Eq.\ (\ref{Eq:B26}). Fig.\ \ref{Fi:6b} also shows the effect of different site energies on $\langle \Delta E(t) \rangle$.

\section{Solution of the equation of motion of the density matrix by linear algebra techniques}\label{Appendix:A}

\begin{figure}[tb]
\begin{center}
\includegraphics[scale=0.60]{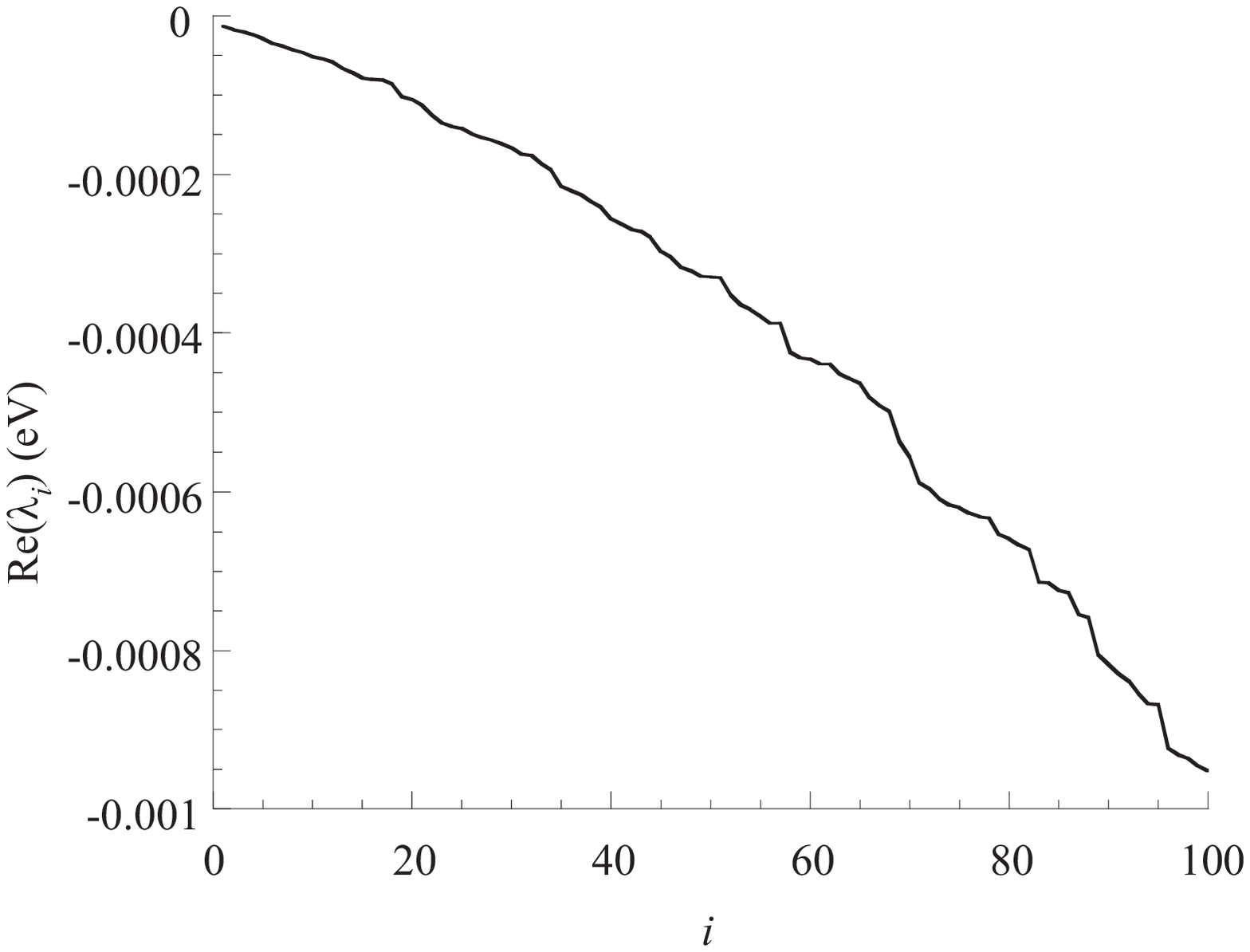}
\end{center}
\caption{$\Re(\lambda_i)$ versus $i$ for the $100$ eigenvalues satisfying $\Re|\lambda_i| \le \Re|\lambda_{100}|$.
$\Re(\lambda_i)$ is off-set from zero by the non-zero radiative
transition rate. } \label{Fi:7}
\end{figure}

As described in Section III, for times greater than $\tau$
the dynamics of the density matrix is determined by eigenmodes of $\textbf{A}$ whose eigenvalues satisfy $\Re|\lambda_i| <
\tau^{-1}$. If only this long-time behavior is
required it is  only necessary to diagonalize
$\textbf{A}$ for a small sub-set of the entire spectrum using
sparse-matrix diagonalization techniques\cite{NAG2}. Fig.\
\ref{Fi:7} shows $\Re(\lambda_i)$ for a lattice of  $25 \times 25$ sites for
$J = 0.00414$ eV. The first $100$ eigenvalues satisfy
$\Re|\lambda_i| < 9.53\times 10^{-4}$ eV, corresponding to
eigenmodes decaying over timescales of  $t > 4.34\times 10^{-12}$ s.

Since $\textbf{A}$ is non-hermitian, the inverse of $\textbf{S}$
(where the columns of  $\textbf{S}$ are the eigenvectors of
$\textbf{A}$) is not equal to its adjoint, and thus its inverse
must be calculated explicitly.  However, if only $N$ out of the
total of $M$ eigenvalues and eigenfunctions have been computed,
 $\textbf{S}$ is a $M \times N$ matrix, and the matrix
equation,
\begin{equation}\label{}
    \textbf{S} \cdot   \textbf{S}^{-1} = \textbf{1}
\end{equation}
is over-determined. Its solution is determined by,
\begin{equation}\label{}
  \textbf{S}^{-1}= \textbf{V} \cdot \left[ \textrm{diag}(1/\lambda_i)\right] \cdot \textbf{U}^{\dagger},
\end{equation}
where $\textbf{V}$ and $\textbf{U}$ are determined by the
singular-value decomposition of  \textbf{S}\cite{num press1,
NAG1},
\begin{equation}\label{}
    \textbf{S}= \textbf{U} \cdot \left[ \textrm{diag}(\lambda_i)\right] \cdot \textbf{V}^{\dagger}.
\end{equation}
(Note that in practice, as Eq.\ (\ref{Eq:16}) indicates, to compute
$\{\rho(t)\}$ only the $m$th column  of $\textbf{S}^{-1}$ is
required.) 

\begin{figure}[tb]
\begin{center}
\includegraphics[scale=0.75]{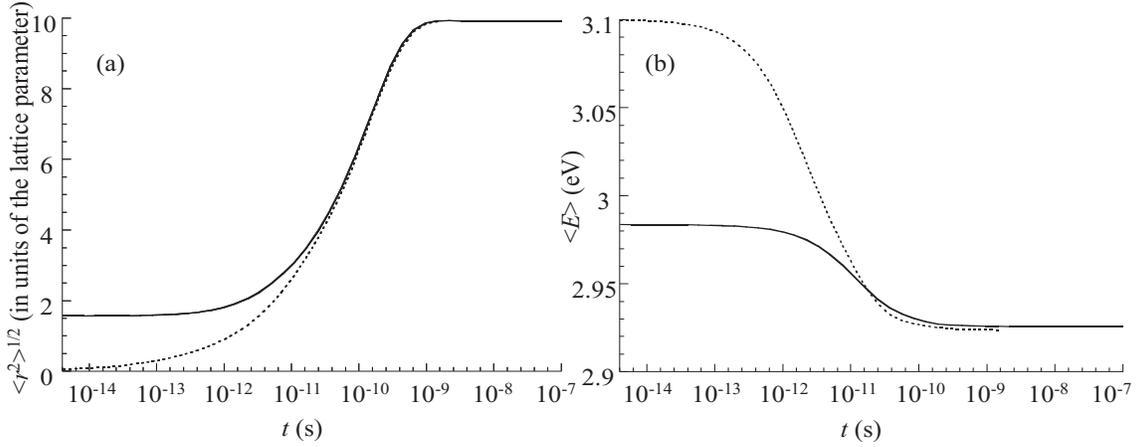}
\end{center}
\caption{Comparison of (a) $\langle r^2 \rangle ^{1/2}$ and (b)
$\langle E \rangle$ obtained by the sparse  matrix diagonalization method
(solid curves) and the Crank-Nicolson method (dotted curves) for a $25 \times 25$ lattice with $J = 0.00414$ eV.}
\label{Fi:8}
\end{figure}

Fig.\ \ref{Fi:8} compares $\langle r^2 \rangle ^{1/2}$
and $\langle E \rangle$ for $25 \times 25$ sites  obtained by the sparse
matrix diagonalization method (using the $100$ highest eigenvalues
shown in Fig.\ \ref{Fi:7}) and the Crank-Nicolson method. As
expected from the eigenvalue spectrum, the results agree for times
$t \gtrsim 10^{-11}$ s.

\begin{acknowledgements}
C. D. P. D. is supported by the EPSRC.
\end{acknowledgements}


\begin{references}

\bibitem[]{email}
Address from 4th September, 2006: Physical and Theoretical Chemistry Laboratory,
University of Oxford, South Parks  Road, Oxford, OX1 3QZ, United Kingdom.

\bibitem{mov1} M. Gr\"unewald, B. Pohlmann, B. Movaghar, and  D. W\"urtz, \textit{Philos. Mag. B} \textbf{49}, 341 (1984)

\bibitem{parson} R. P. Parson and R. Kopelman, \textit{J. Chem. Phys.} \textbf{82}, 3692 (1985)

\bibitem{mov2} B. Movaghar, M. Gr\"unewald, B. Ries, H. B\"assler,  and  D. W\"urtz, \textit{Phys. Rev. B} \textbf{33}, 5545 (1986)

\bibitem{mov3} B. Movaghar,  B. Ries,  and  M. Gr\"unewald, \textit{Phys. Rev. B} \textbf{34}, 5574 (1986)

\bibitem{meskers} S. C. J. Meskers, J. H\"uber, M. Oestreich, and H. B\"assler, \textit{J. Phys. Chem. B} \textbf{105}, 9139 (2001)

\bibitem{amerongen} H. van Amerongen, L. Valkunas, and R. van Grondelle, \textit{Photosynthetic Excitons}, World Scientific, Singapore (2000)

\bibitem{grondelle} R. van Grondelle and V. I. Novoderezhkin, \textit{Phys. Chem. Chem. Phys.} \textbf{8}, 793 (2006)

\bibitem{anderson} P. W. Anderson, \textit{Phys. Rev.} \textbf{109}, 1492 (1958)

\bibitem{bassler06} H. B\"assler, \textit{Nature Physics} \textbf{2}, 15 (2006)

\bibitem{dubin06} F. Dubin, R. Melet, T. Barisien, R. Grousson, L. Legrand, M. Schott, and V. Voliotis,
\textit{Nature Physics}, \textbf{2}, 32 (2006)

\bibitem{milota04} F. Milota, J. Sperling, V. Szoecs, A. Tortschanoff, and H. F. Kauffmann, \textit{J. Chem. Phys.}
\textbf{120}, 9870 (2004)

\bibitem{may} V. May and O. K\"uhn,
\emph{Charge and Energy Transfer Dynamics in Molecular Systems}, Wiley-VCH,  Weinheim (2004)

\bibitem{reineker} P. Reineker,
\emph{Exciton Dynamics in Molecular Crystals and Aggregates}, Springer Tracts in Modern Physics, vol 94,  Berlin (1982)

\bibitem{silinsh} E. A. Silinsh and V. Capek, \emph{Organic Molecular Crystals}, AIP Press, New York (1994)

\bibitem{num press2} W. H. Press, S. A. Teukolsky, W. T. Vetterling, and B. P. Flannery,
\emph{Numerical Recipes in Fortran 77}, Vol. 1, 2nd. edition, p. 838, Cambridge University Press,  Cambridge (1992)

\bibitem{NAG3} The implicit-in-time solutions were found by solving the complex tridiagonal matrix with the
NAG routine \emph{F07CNF}.

\bibitem{num press3} W. H. Press, S. A. Teukolsky, W. T. Vetterling, and B. P. Flannery,
\emph{Numerical Recipes in Fortran 77}, Vol. 1, 2nd. edition, p. 847, Cambridge University Press,  Cambridge (1992)

\bibitem{footnote1} Since exciton dephasing occurs via the collision of an exciton with a defect, the dephasing rate is in principle proportional to $J$. In this paper we ignore this dependency.

\bibitem{num press4} Using the \textit{GASDEV} routine from W. H. Press, S. A. Teukolsky, W. T. Vetterling, and B. P. Flannery,
\emph{Numerical Recipes in Fortran 77}, Vol. 1, 2nd. edition, p. 280, Cambridge University Press,  Cambridge (1992)

\bibitem{NAG2} The sparse-diagonalization of \textbf{A} was achieved by the ARPACK routines, \emph{F12ANF},
 \emph{F12ARF}, \emph{F12APF}, \emph{F12ASF}, and \emph{F12AQF} supplied by NAG.

\bibitem{num press1} W. H. Press, S. A. Teukolsky, W. T. Vetterling, and B. P. Flannery,
\emph{Numerical Recipes in Fortran 77}, Vol. 1, 2nd. edition, p. 57, Cambridge University Press,  Cambridge (1992)

\bibitem{NAG1} The singular-value decomposition of \textbf{S} was obtained by the LAPACK routine \emph{F08KPF}
 supplied by NAG.


\end{references}
\end{document}